\documentclass[twoside]{LCWS11}
\usepackage[latin1]{inputenc}
\usepackage[pdftex]{graphicx}
\usepackage{wrapfig,rotating}
\usepackage{amssymb,amsmath,array}
\usepackage{cite}
\usepackage{units}
\usepackage{heppennames2}
\usepackage{subfig}
\DeclareGraphicsExtensions{.pdf, .png}
\begin{document}

\newcommand{\epem}{\ensuremath{\mathrm{\Pep\Pem}}\xspace}
\newcommand{\abinv}{\ensuremath{\mathrm{ab}^{-1}}\xspace}
\newcommand{\pT}{\ensuremath{p_\mathrm{T}}\xspace}

\newcommand{\gghadrons}{\ensuremath{\upgamma\upgamma \rightarrow \mathrm{hadrons}}\xspace}
\newcommand{\clicsid}{CLIC\_SiD\xspace}
\newcommand{\micron}{\ensuremath{\upmu\mathrm{m}}}
\newcommand{\radlen}{\ensuremath{X_{0}}\xspace}
\newcommand{\radlenfrac}{\ensuremath{X/X_{0}}\xspace}
\newcommand{\nuclen}{\ensuremath{\lambda_{\mathrm{I}}}\xspace}
\newcommand{\degrees}{\ensuremath{^{\circ}}\xspace}
\newcommand{\rmlad}{\ensuremath{_{\mathrm{ladder}}}}
\newcommand{\mm}[1]{\ensuremath{_{\mathrm{#1}}~\mathrm{[mm]}}}
\newcommand{\mic}[1]{\ensuremath{_{\mathrm{#1}}~\mathrm{[\micron]}}}
\newcommand{\mumu}{\ensuremath{\upmu\upmu}\xspace}
\newcommand{\nuenuebar}{\ensuremath{\PGne\PAGne}\xspace} 
\newcommand{\mpmm}{\ensuremath{\PGmp\PGmm}\xspace}  
\newcommand{\nunubar}{\ensuremath{\PGn\PAGn}\xspace}   
\newcommand{\tptm}{\ensuremath{\PGtp\PGtm}\xspace} 
\newcommand{\gamgam}{\ensuremath{\upgamma\upgamma}\xspace}
\newcommand{\ww}{\ensuremath{\PWp\PWm}\xspace} 
\newcommand{\zz}{\ensuremath{\PZz\PZz}\xspace} 
\newcommand{\wwz}{\ensuremath{\PWp\PWm\PZz}\xspace} 
\newcommand{\zzz}{\ensuremath{\PZz\PZz\PZz}\xspace} 
\newcommand{\zhsm}{\ensuremath{\PH\PZz}\xspace}  
\newcommand{\hbb}{\ensuremath{\PH\to b\bar{b}}\xspace}
\newcommand{\hcc}{\ensuremath{\PH\to c\bar{c}}\xspace}
\newcommand{\hmumu}{\ensuremath{\PH\to \mumu}\xspace}
\newcommand{\guineapig}{\textsc{GuineaPig}\xspace}
\newcommand{\mokka}{\textsc{Mokka}\xspace}
\newcommand{\marlin}{\textsc{Marlin}\xspace}
\newcommand{\geant}{\textsc{Geant4}\xspace}
\newcommand{\slic}{\textsc{SLIC}\xspace}
\newcommand{\lcsim}{\texttt{org.lcsim}\xspace}
\newcommand{\pythia}{\textsc{PYTHIA}\xspace}
\newcommand{\whizard}{\textsc{WHIZARD}\xspace}
\newcommand{\pandora}{\textsc{PandoraPFA}\xspace}
\newcommand{\fastjet}{\textsc{FastJet}\xspace}
\newcommand{\tmva}{TMVA\xspace}
\newcommand{\roofit}{RooFit\xspace}

\title{Measurement of the Cross Section Times Branching Ratio of Light Higgs Decays at CLIC}
\author{Christian Grefe$^1$ Tomas Lastovicka$^2$ Jan Strube$^1$ Frederic Teubert$^1$ Blai Pie Valls$^3$
\vspace{.3cm}\\
1- CERN \\
Meyrin - Switzerland
\vspace{.1cm}\\
2- Institute of Physics, Academy of Sciences of the Czech Republic\\
Prague - Czech Republic
\vspace{.1cm}\\
3- Faculty of Physics, Universitat de Barcelona \\
Barcelona - Spain\\
}

\maketitle

\begin{abstract}
The investigation of the properties of a Higgs boson, especially a test of the predicted linear dependence of the branching ratios on the mass of the final state, is currently one of the most compelling arguments for building a linear collider. We demonstrate that the large Higgs boson production cross section at a \unit[3]{TeV} CLIC machine allows for a precision measurement of the Higgs branching ratios. The cross section times branching ratio of the decays \hbb, \hcc and \hmumu can be measured with a statistical uncertainty of 0.22\%,  3.2\% and 15\%, respectively.
\end{abstract}

The Higgs mechanism of electroweak symmetry breaking predicts the existence of a fundamental spin-0 particle that has not yet been discovered. The LHC will most probably answer the question about its existence by 2013. The Standard Model predicts a linear dependence of the Higgs branching ratios on the mass of the final state, but non-standard couplings could alter this relation. The detailed exploration of the Higgs sector will be instrumental to our understanding of the fundamental interactions. The LHC will be of limited utility in this exploration.

The compact linear collider (CLIC) is a proposed \epem collider with a maximum center-of-momentum energy $\sqrt{s} = \unit[3]{TeV}$, based on a two-beam acceleration scheme~\cite{CLICacceleratorCDR}. We present here the analysis of the branching ratios \hbb, \hcc~\cite{lcd:2011-036} and \hmumu~\cite{lcd:grefeHmumu2011} at such a machine. The studies are based on fully simulated samples in the \clicsid~\cite{lcd:grefemuennich2011} detector concept and take into account the main beam-related background.

\section{The \clicsid Detector Model}
\label{sec:DetectorModel}
The detector that is used in the full simulation of samples is based on the SiD concept~\cite{Aihara:2009ad} and has been adapted\cite{lcd:grefemuennich2011} to the specific detector requirements at CLIC. It is designed for particle flow calorimetry using highly granular calorimeters.

A superconducting solenoid with an inner radius of \unit[2.9]{m} provides a central magnetic field of \unit[5]{T}. The calorimeters are placed inside the coil and consist of a 30 layer tungsten-silicon electromagnetic calorimeter with \unit[$3.5\times3.5$]{mm$^2$} segmentation, followed by a tungsten-scintillator hadronic calorimeter with 75 layers in the barrel region and a steel-scintillator hadronic calorimeter with 60 layers in the endcaps. The read-out cell size in the hadronic calorimeters is \unit[$30\times30$]{mm$^2$}. The iron return yoke outside of the coil is instrumented with 9 double RPC layers with \unit[$30\times30$]{mm$^2$} read-out cells for muon identification.

The silicon-only tracking system consists of 5 \unit[$20\times20$]{$\micron^2$} pixel layers followed by 5 strip layers with a pitch of \unit[25]{\micron}, a read-out pitch of \unit[50]{\micron} and a length of \unit[92]{mm} in the barrel region. The tracking system in the endcap consists of 5 strip disks with similar pitch and a stereo angle of 12\degrees, complemented by 7 pixelated disks in the vertex and far-forward region at lower radii with pixel sizes of of \unit[$20\times20$]{$\micron^2$}.

\section{Software and Data Samples}
\label{sec:Samples}
The physical processes are produced with the Whizard~\cite{Kilian:2007gr,whizard2} event generator, with fragmentation and hadronization by \pythia~\cite{Sjostrand2006}. The events are simulated in the CLIC\_SID detector model using SLIC, a thin wrapper around GEANT4~\cite{Allison:2006ve}. The event reconstruction is handled by the lcsim and slicPandora packages, the LCFI package is used for flavor tagging. The assumed luminosity of the analyses is 2 \abinv, corresponding to about 4 years of data taking at nominal conditions.

The predominant production channel of Higgs bosons at a \unit[3]{TeV} linear collider is the WW fusion channel $\epem \to h \nu \nu$.
The main backgrounds are from events with two Z bosons, where one Z decays to the signal final states, and the other decays to neutrinos, or decays to a fermion pair that does not enter the fiducial volume of the detector. The beam configuration at a \unit[3]{TeV} CLIC machine produces 3.2 \gghadrons events per bunch crossing on average. With a spacing of \unit[0.5]{ns} between bunches, these necessarily pile up in the subdetectors, which have integration times of \unit[10]{ns}, except for the barrel hadronic calorimeter, which has an integration time of \unit[100]{ns}. To take into account the effect of this background on the measurement, a fully simulated sample of events from \gghadrons corresponding to 60 bunch crossings is mixed with each physics event for the analysis of the Higgs decaying to b and c quarks. In the \hmumu analysis, only the signal sample was mixed with events from \gghadrons background.
Table~\ref{tab:samples} lists the physics processes that were taken into account in the analyses, together with their cross section and the number of simulated events.
\begin{table}
 \centering
\begin{tabular}{p{6cm} r r l}
\hline \hline
Process & $\sigma$ [fb] & $N_{\mathrm{events}}$ & Short label \\
\hline
$\epem \to \PH \nuenuebar$; \hmumu    & 0.120  &   21000 & \hmumu    \\
$\epem \to \PH \nuenuebar$; \hbb &   285  &   45000 & \hbb  \\
$\epem \to \PH \nuenuebar$; \hcc &    15  &  130000 & \hcc \\
\hline
$\epem \to \mpmm \nunubar$                       & $132^{*}$          & 5000000 & $\mpmm \nunubar$    \\
$\epem \to \mpmm \epem$                          & $346^{*}$          & 1350000 & $\mpmm \epem$       \\
$\epem \to \mpmm$                                &  $12^{*}$ 	   &   10000 & $\mpmm$             \\
$\epem \to \tptm$                                & $250^{*}$         &  100000 & $\tptm$             \\
$\epem \to \tptm \nunubar$                       & $125^{*}$         &  100000 & $\tptm \nunubar$    \\
$\epem \to qq$								     & 3100	&   96000 & $qq$				  \\
$\epem \to qq\nunubar$						     & 1300	&  170000 & $qq\nu\nu$	      \\
$\epem \to qq\epem$							     & 3300	&   90000 & $qq\epem$			  \\
$\epem \to q q e\nu$						     & 5300	&   91000 & $qq e \nu$          \\
$\gamgam \to \mpmm$ (generator level only)       & $20000^{*}$        & 1000000 & $\gamgam \to \mpmm$ \\
\hline \hline
\end{tabular}
 \caption{List of processes considered for this analysis with their respective cross section $\sigma$ and the number of simulated events $N_{\mathrm{events}}$. The cross section takes into account the CLIC luminosity spectrum. Cross sections marked with * include a cut on the invariant mass of the muon pair to lie between 100 and 140 GeV.}
 \label{tab:samples}
\end{table}

\section{Measurement of \hbb and \hcc}
\subsection{Event Selection}
The main background of the measurement of the decays \hbb and \hcc is from two-jet processes $\epem\to qq\nu\nu$, due to their large cross section, and from processes with two measured jets and additional particles that escape detection.

The FastJet package is used to cluster the events into two jets. The LCFI flavor tagging package finds secondary vertices in each jet and uses them in a neural network to distinguish b-, c-, and light quark jets. The invariant mass of the jet pair is the major discriminant between decays of Higgs and of Z bosons. It is used in a second neural network, together with the output of the flavor-tagging network and the following variables:
\begin{itemize}
\item the maximum of the absolute values of jet pseudorapidities
\item the sum of the remaining LCFI jet flavor tag values, i.e., c(udsb), c(b)-tags and b(uds)-tag
\item $R_{\eta\phi}$, the distance of jets in the $\eta-\phi$ plane
\item the sum of jet energies
\item the total number of leptons in an event
\item the total number of photons in an event
\item the acoplanarity of the jets
\end{itemize}

\begin{figure}
	\subfloat[\hbb]{\includegraphics[width=.5\textwidth]{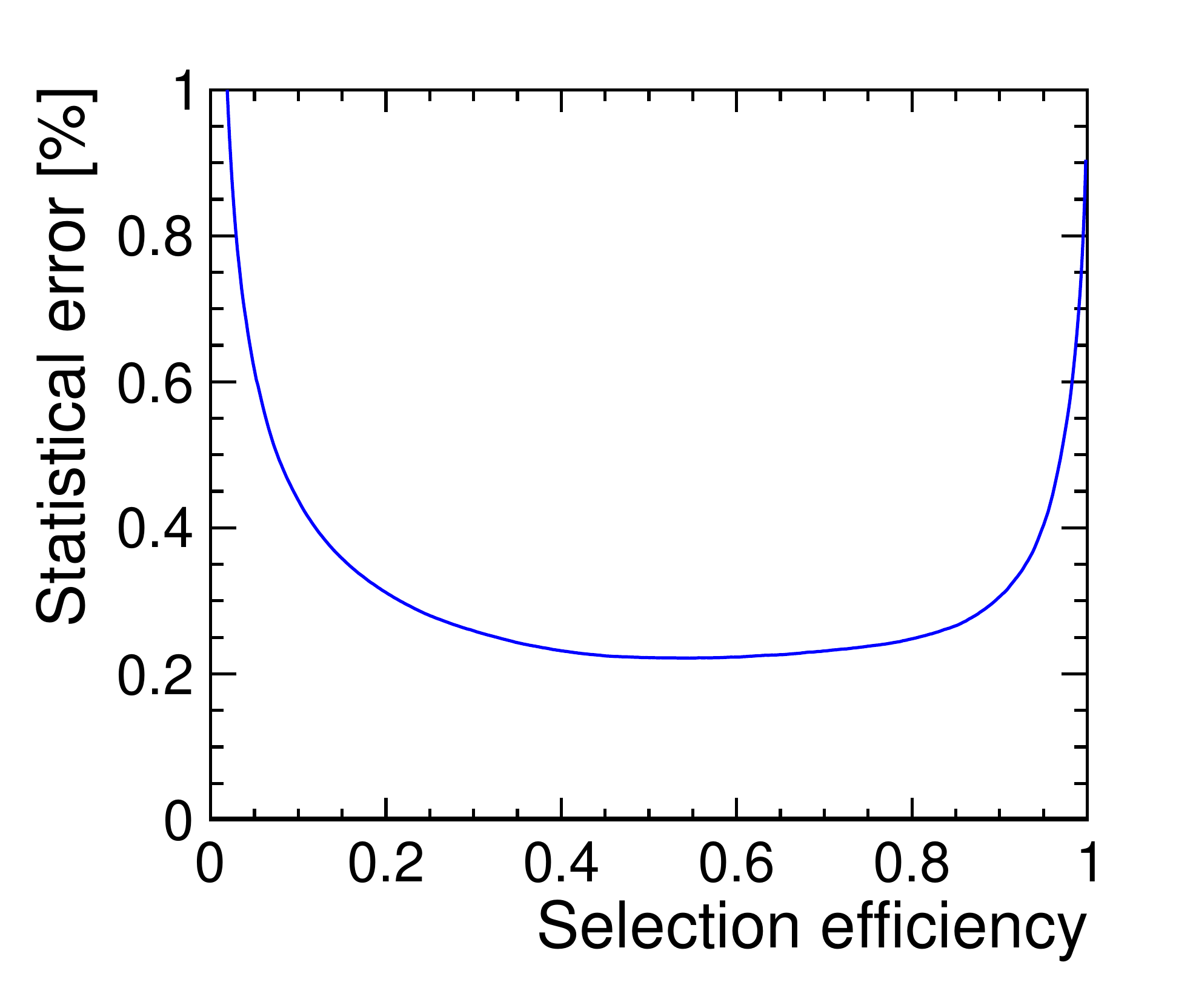}} \hfill
	\subfloat[\hcc]{\includegraphics[width=.5\textwidth]{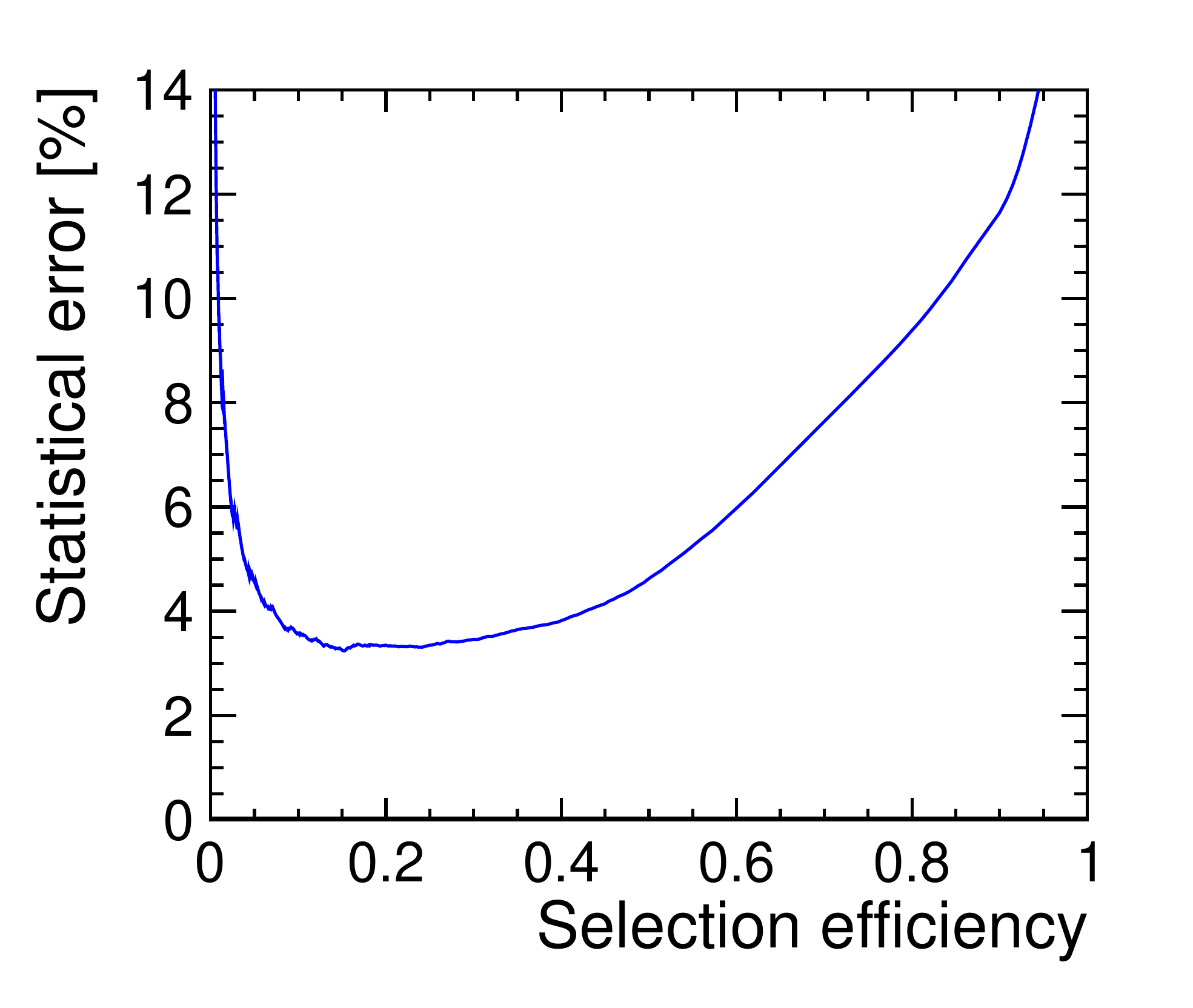}}
	\caption{Statistical uncertainty of the measurement of cross section times branching ratio versus selection efficiency of the neural network. Left: The neural network was trained to identify \hbb decays from di-jet backgrounds including \hcc. Right: The neural network was trained on \hcc as signal and di-jets including \hbb as background.}
	\label{fig:bcSelectionEfficiency}
\end{figure}

The neural network selection efficiency $S/(S+B)$ versus the statistical uncertainty $\sqrt{S+B}/S$ on the measurement of the number of signal events $S$ and background events $B$ is shown in Figure~\ref{fig:bcSelectionEfficiency} for the two neural networks that were trained on \hbb and \hcc as signal, respectively. The optimal selection is at the local minimum of the curve, at a selection efficiency of 55\% for \hbb with a sample purity of 65\%, corresponding to a statistical uncertainty of 0.22\%. The optimal selection for \hcc has an efficiency of 15\%, corresponding to a sample purity of 24\% and a measurement uncertainty of 3.2\%. These values reflect the fact that b-jets can be distinguished from c-jets with high purity, while incompletely reconstructed b-jets make up a large fraction of the background to c-jet selection, making the analysis more challenging. Using the output of the reconstruction algorithms in neural networks leads to the minimal statistical uncertainty on the measurement at the cost of an increased dependence on systematic effects. We assume that with sufficient experience at the running machine, the systematic variations are well enough understood so that the systematic uncertainties are comparable to the statistical uncertainties.

\section{Measurement of \hmumu}
 The measurement of the rare decay \hmumu requires high luminosity operation and sets stringent limits on the momentum resolution of the tracking detectors. The branching ratio of the decay of a Standard Model Higgs boson to a pair of muons is important as the lower end of the accessible decays and defines the endpoint of the test of the predicted linear dependence of the branching ratios to the mass of the final state particles.

\section{Event Selection}
\label{sec:EventSelection}
\begin{wrapfigure}{r}{.5\linewidth}
 \centering
 \includegraphics[width=\linewidth]{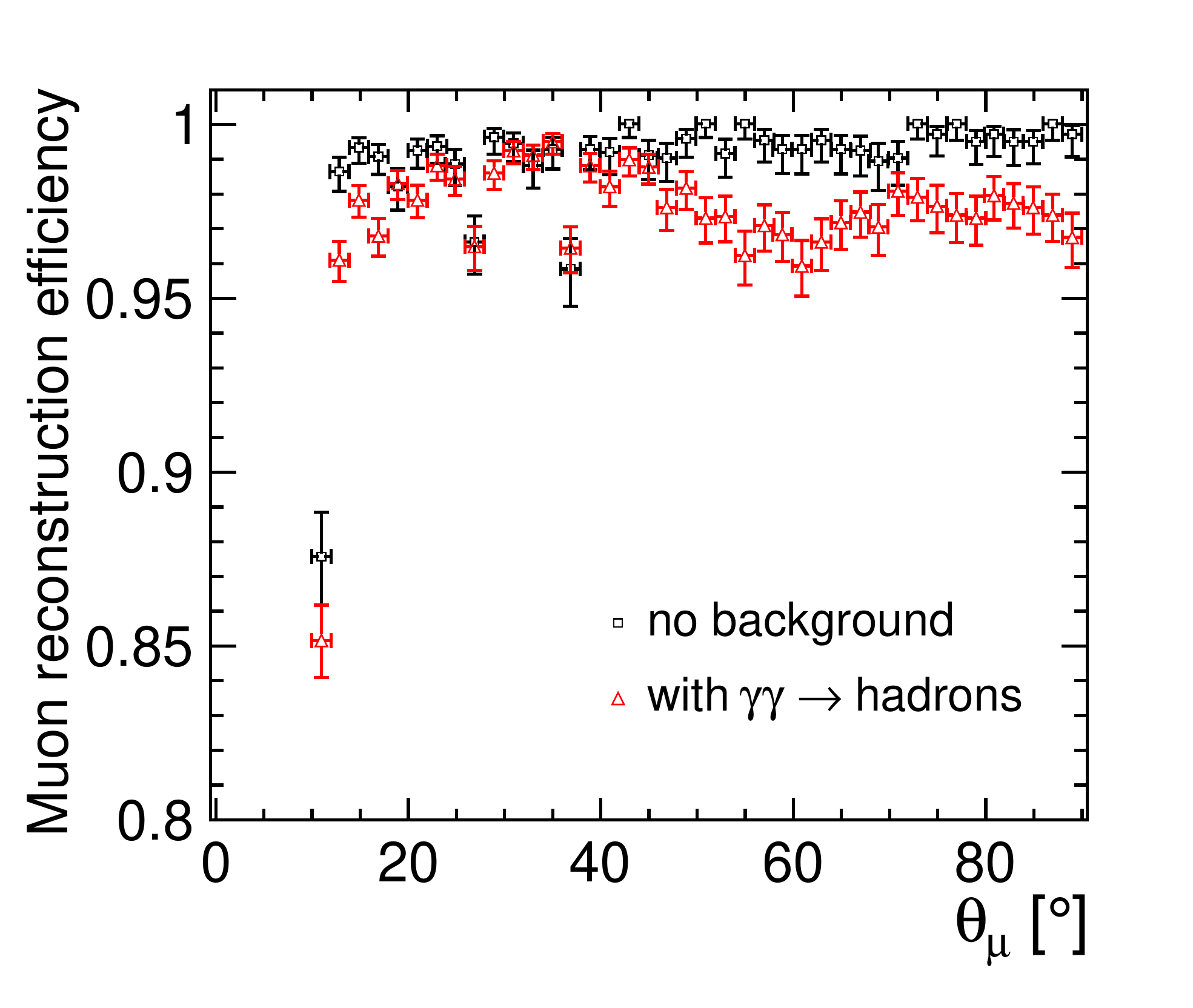}
\caption{Muon reconstruction efficiency for the signal sample with and without \gghadrons pile-up}
\label{fig:background_evis}
\end{wrapfigure}
The muon reconstruction efficiency in the signal sample is shown in Figure~\ref{fig:background_evis}. The small deterioration of the muon reconstruction efficiency due to beam-induced background from \gghadrons is evident. The average muon reconstruction efficiency for polar angles greater than 10\degrees is 98.4\% with this background compared to 99.6\% without. The total reconstruction efficiency of the signal sample, requiring two reconstructed muons with an invariant mass between \unit[105]{GeV} and \unit[135]{GeV} is 72\% in the presence of background.
The events are selected by requiring two reconstructed muons, each with a transverse momentum of at least \unit[5]{GeV}. In case there are more than two muons reconstructed, the two most energetic ones are used. In this note, the most energetic muon is referred to as $\upmu_1$ and the second most energetic muon is referred to as $\upmu_2$. In addition, the invariant mass of the two muons $M(\mumu)$ is required to lie between \unit[105]{GeV} and \unit[135]{GeV}.

The event selection is done using the boosted decision tree classifier implemented in \tmva~\cite{TMVA:2010}. The $\mpmm$, $\tptm$ and $\tptm \nunubar$ samples are not used in the training of the BDT, but are effectively removed by the classifier nevertheless.

The variables used for the event selection by the BDT are:
\begin{itemize}
 \item The visible energy excluding the two reconstructed muons $E_{\mathrm{vis}}$.
 \item The scalar sum of the transverse momenta of the two muons $\pT(\upmu_1) + \pT(\upmu_2)$.
 \item The helicity angle $\cos\theta^*(\mumu) = \frac{\vec{p}'(\upmu_1) \cdot \vec{p}(\mumu)}{|\vec{p}'(\upmu_1)| \cdot |\vec{p}(\mumu)|}$, where $\vec{p}'$ is the momentum in the rest frame of the di-muon system. Since the two muons are back-to-back in the rest frame of the di-muon system there is no additional information to be gained from calculating a similar angle for $\upmu_2$.
 \item The relativistic velocity of the di-muon system $\upbeta(\mumu)$, where $\upbeta = \frac{v}{c}$.
 \item The transverse momentum of the di-muon system $\pT(\mumu)$.
 \item The polar angle of the di-muon system $\theta(\mumu)$.
\end{itemize}

\begin{wrapfigure}{r}{0.5\columnwidth}
\centering
\includegraphics[width=\linewidth]{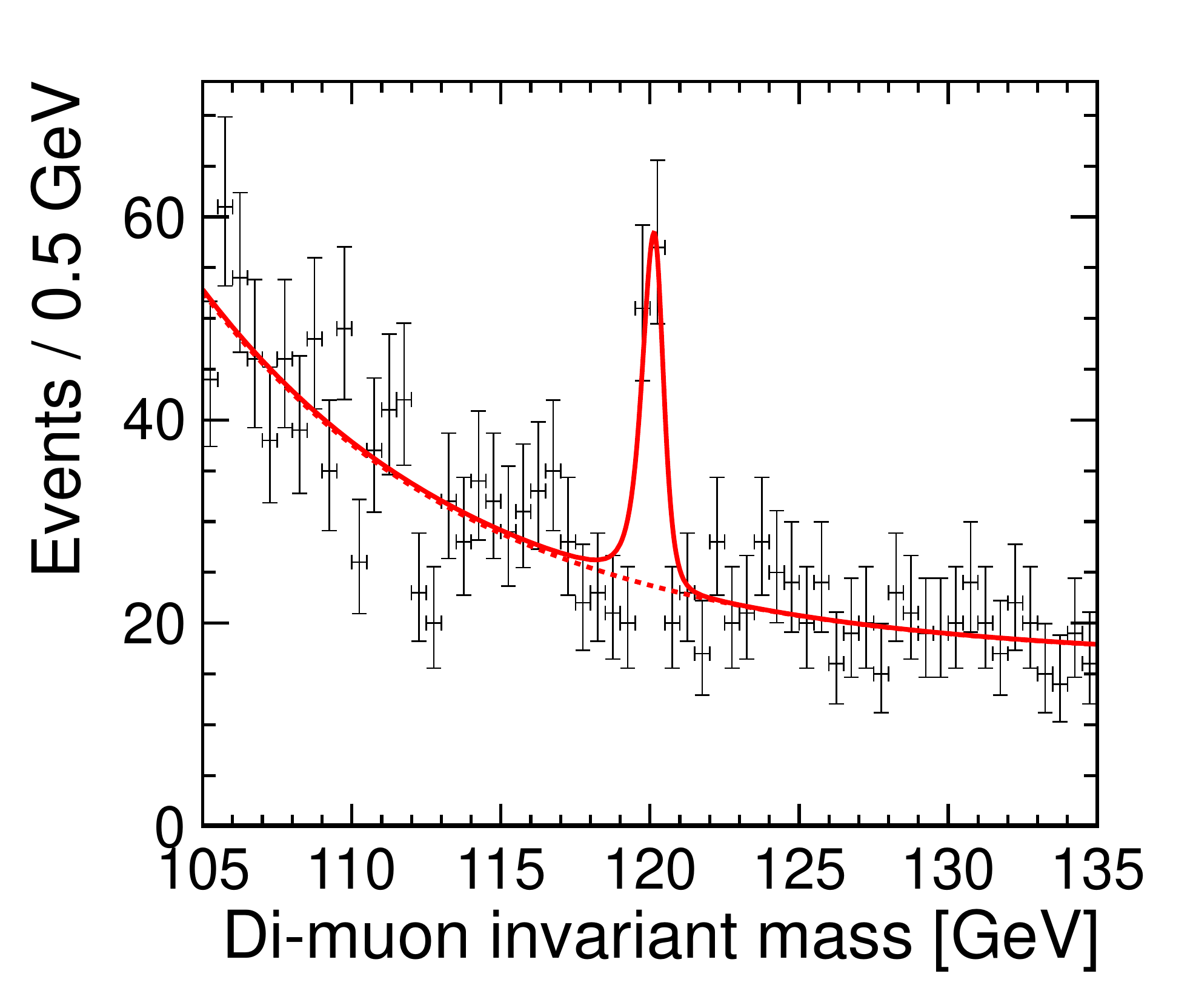}
\caption{Maximum Likelihood fit of the Higgs mass in the data sample after selection cuts}\label{fig:mumu_massFit}
\end{wrapfigure}

The most powerful variable is the visible energy whenever there is an electron within the detector acceptance. Otherwise the background can be rejected by the transverse momentum of the di-muon system or the sum of the two individual transverse momenta.
Figure~\ref{fig:mumu_massFit} clearly shows the Higgs peak in the invariant mass distribution after the event selection.
The background from $\epem \to \epem\mumu$ events, is effectively reduced by forward electron tagging. While the forward calorimeters were not part of the full detector simulation, assuming a tagging efficiency of 95\% down to an angle of 40 mrad for electrons of several hundred GeV to over one TeV is a conservative estimate, even in the presence of \gghadrons background. It is found that Bhabha events prevent further rejection of this background at lower angles. The results quoted are based on a ad-hoc rejection of 95\% of the electrons in the Luminosity Calorimeter. 

\subsection{Invariant Mass Fit}
\label{sec:MassFit}	
The distribution of the invariant mass in the \hmumu sample has a tail towards lower masses because of final state radiation. The shape can be described best by two half Gaussian distributions with an exponential tail. Together with the mean value this results in five free parameters in the fitted function, which can be written as
\begin{equation*}
 f(x) = n \left\{ \begin{array}{rl}
                 e^{\frac{x - m_0}{2\sigma_L^2 + \alpha_L(x - m_0)^2}} &\mbox{, $x \leq m_0$} \\
                 e^{\frac{x - m_0}{2\sigma_R^2 + \alpha_R(x - m_0)^2}} &\mbox{, $x > m_0$}
                \end{array} \right . ,
 \label{eq:MassFit_hmumu}
\end{equation*}
where $m_0$ is the mean of both Gaussian distributions, $\sigma_L$ and $\sigma_R$ are the widths, and $\alpha_L$ and $\alpha_R$ are the tail parameters of the left and the right Gaussian distribution, respectively; $n$ is a normalization parameter.
The background is well described by an exponential parameterization, obtained from a background-only sample.

The number of signal events is obtained from a maximum likelihood fit to the sample containing signal plus background after the event selection.

\subsection{Study of the Momentum Resolution}
\label{sec:MomentumResolution}
\begin{wraptable}{r}{.5\linewidth}
 \centering
\begin{tabular}{r r r}
\hline\hline
$\sigma(\Delta\pT)/\pT^2$    & $\sigma(\Delta M(\mumu))$ & Stat. \\
& & uncertainty \\
\hline
\unit[$10^{-3}$]{GeV$^{-1}$} & \unit[6.5]{GeV}           &   -               \\
\unit[$10^{-4}$]{GeV$^{-1}$} & \unit[0.70]{GeV}          & 34.3\%            \\
\unit[$10^{-5}$]{GeV$^{-1}$} & \unit[0.068]{GeV}         & 18.2\%            \\
\unit[$10^{-6}$]{GeV$^{-1}$} & \unit[0.022]{GeV}         & 16.0\%            \\
\hline\hline
\end{tabular}
 \caption{Summary of the results for the $\PSh \to \mpmm$ branching ratio measurement using fast simulation samples with different momentum resolutions $\sigma(\Delta\pT)/\pT^2$ assuming an integrated luminosity of $\unit[2]{\abinv}$. Given is the corresponding invariant mass resolution $\sigma(\Delta M(\mumu))$ and the resulting statistical uncertainty of the $\sigma_{\PSh\nuenuebar} \times \mathrm{BR}_{\PSh \to \mpmm}$ measurement. The results do not include reduction of the $\epem \to \epem \mumu$ background using ad-hoc electron tagging.}
 \label{tab:resultsMomentumResolution}
\end{wraptable}
The ability to measure the decay \hmumu depends crucially on the momentum resolution of the tracking detectors. In a fast simulation study, different values for the momentum resolution were assumed, before the sample was fit as described in the previous section.
The results are shown in Table~\ref{tab:resultsMomentumResolution}. It is found that an average resolution of $5\times\unit[10^{-5}]{GeV^{-1}}$ or better is required in order for the momentum resolution not to be the dominant uncertainty contribution in a \unit[2]{\abinv} measurement of the decay \hmumu. The average muon momentum resolution of the fully simulated sample is $4\times\unit[10^{-5}]{GeV^{-1}}$ corresponding to a statistical uncertainty of 23\% without forward electron tagging.

\subsection{Results}
\label{sec:result}
We have demonstrated the feasibility of measuring the branching ratios of a 120 GeV Higgs boson at a 3 TeV CLIC with high precision.
For the measurement of Higgs decays to quarks, 0.22\% and 3.2\% statistical uncertainty can be achieved for the decays \hbb and \hcc, respectively. This includes the effect of background from \gghadrons on the flavor tagging.

For the rare decay \hmumu, the cross section times branching ratio can be measured to a precision of 15\% if the background from $\epem\to\mpmm$ can be reduced using tagging of electrons down to an angle of 40 mrad with an efficiency of 95\%, and the average momentum resolution is not worse than $5\times 10^{-5}$. The effect of background from \gghadrons has been taken into account.
From experience of the LEP experiments one can assume that the systematic uncertainties related to detector effects are of the order of 1\% or less. For the measurement of $\sigma_{\PZz \to \mpmm}$ at LEP the systematic uncertainty was between 0.1 and 0.4\%, depending on the experiment. In summary, one can expect that the systematic uncertainty of this analysis will be negligible compared to the statistical uncertainty.

The expected uncertainty of the peak luminosity is currently being studied but is estimated to be around 1\% or less.

\bibliographystyle{hplain}
\bibliography{cdrbibliography}




\end{document}